\documentclass[conference]{IEEEtran}
\IEEEoverridecommandlockouts

\usepackage{graphicx}
\usepackage{amsmath, amssymb}
\usepackage{siunitx}
\usepackage{booktabs}
\usepackage{multirow}
\usepackage{hyperref}
\usepackage{cite}
\usepackage{listings}
\usepackage{xcolor}
\usepackage{array}
\usepackage{balance}
\usepackage{booktabs}
\usepackage{subcaption}
\usepackage[percent]{overpic}

\hypersetup{
    colorlinks=true,
    linkcolor=black,
    citecolor=black,
    urlcolor=black
}

\title{Reinforcement Learning for Vehicle-to-Grid Voltage Regulation: Single-Hub to Multi-Hub Coordination with Battery-Aware Constraints}

\author{\IEEEauthorblockN{Jingbo Wang, \textit{Graduate Student Member, IEEE,} Roshni Anna Jacob, \textit{Member, IEEE,} \\ Harshal D. Kaushik, \textit{Member, IEEE,}  Jie Zhang, \textit{Senior Member, IEEE}}
    \IEEEauthorblockA{The University of Texas at Dallas,  Richardson, Texas 75080, USA \\
    \ Email: jiezhang@utdallas.edu}
}

\begin{document}
\maketitle

\begin{abstract}
This paper presents a Vehicle-to-Grid (V2G) coordination framework using reinforcement learning (RL). {An intelligent control strategy based on the soft actor-critic algorithm is developed for voltage regulation through single and multi-hub charging systems while respecting realistic fleet constraints. A two-phase training approach integrates stability-focused learning with battery-aware deployment to ensure practical feasibility. Simulation studies on the IEEE 34-bus system validate the framework against a standard Volt-Var/Volt-Watt droop controller. Results indicate that the RL agent achieves performance comparable to the baseline control strategy in nominal scenarios. Under aggressive overloading, it provides robust voltage recovery (within 10\% of the baseline) while prioritizing fleet availability and state-of-charge preservation, demonstrating the viability of constraint-aware learning for critical grid services.}
\end{abstract}


\section{Introduction}

The rapid proliferation of electric vehicles (EVs) is transforming distribution grids, introducing voltage challenges while opening possibilities for Vehicle-to-Grid (V2G) services. Bidirectional V2G systems enable EVs to function as distributed energy resources (DERs), modulating active and reactive power to support voltage regulation~\cite{alamgir2025comprehensive}. In this context, we focus on EV ``fleets'' (e.g., logistics or ride-sharing entities) where a central operator controls charging and discharging actions across either a single hub or a multi-hub system. However, effectively harnessing this flexibility requires sophisticated control frameworks that integrate network constraints with realistic battery dynamics.

Traditional voltage regulation relies on legacy devices with limited actuation speeds and discrete settings, rendering them inadequate for rapid fluctuations. In contrast, inverter-based DERs are increasingly gaining attention in this regard, as they can be controlled to offer continuous response with sub-second response times. In parallel, reinforcement learning (RL) has emerged as a promising solution for developing intelligent control strategies. RL has been applied to various critical grid operations, from real-time outage management~\cite{jacob2024real,chen2023learning} to distribution grid voltage control~\cite{yang2019two,hossain2022soft}. While these studies demonstrate RL's capabilities, existing works on voltage regulation often model controllable resources with static capacity limits, neglecting temporal or state-dependent flexibility constraints inherent in V2G systems~\cite{nematshahi2023deep, cui2022decentralized}.

Despite advancements in RL-based voltage control, several critical gaps remain in addressing realistic V2G system constraints. Existing studies predominantly focus on single aggregators or homogeneous fleets, leaving the coordination of spatially distributed charging hubs (within a feeder) operating under unified control policies largely underexplored. Moreover, most distribution network control models assume static battery capacity bounds and neglect key factors such as state of charge (SOC) or state of health (SOH) dynamics, as well as continuously derated power envelopes that arise with real-world EV deployment. While existing studies focus on either single-hub or multi-hub scenarios in isolation, the scalability challenges of expanding from single-hub to multi-hub V2G systems are not adequately addressed. The aggregated behavior of EV fleets is influenced by heterogeneous factors SOC, SOH, availability patterns, and nonlinear battery constraints, which must be integrated into control frameworks for practical deployment~\cite{ledro2022influence, xie2025reinforcement}. 

Building on prior work in EV infrastructure planning that addressed optimal charging station deployment considering hosting capacity and voltage constraints~\cite{2025_PESGM, 2025_iScience}, this paper shifts focus to operation, aiming to utilize EV flexibility to improve voltage regulation. To address these gaps, this paper proposes a V2G control architecture that enables both single-hub and multi-hub coordination, employing RL-based intelligent control to dispatch active and reactive power across multiple geographically distributed hubs. A fleet-aware power mapping module translates hub-level power signals into battery-level actions by incorporating power conversion, inverter efficiency, and SOC/SOH-dependent bounds.

The remainder of this paper is organized as follows: Section \ref{sec:methodology} presents the methodology, including the system overview, distribution network hub, EV fleet models, and the RL framework. Section \ref{sec:setup} describes the case study setup and RL agent training environment. {Section \ref{sec:results} presents the simulation results, providing a comparative analysis against industry-standard Volt-Var/Volt-Watt local control strategies to validate the proposed framework's efficacy in voltage regulation while strictly adhering to fleet constraints.} Finally, Section \ref{sec:conclusion} concludes the paper with key findings and future research directions. %Section \ref{sec:results} presents the results for both single-hub and multi-hub voltage regulation, followed by performance analysis.

\section{Methodology}
\label{sec:methodology}

This section presents the methodological framework for voltage regulation through coordinated V2G control using RL, while integrating power flow simulation and realistic EV fleet modeling with the learning-based control.

\subsection{System Overview and Hub Model}

The system under consideration comprises a radial distribution feeder integrated with V2G hubs at select buses. Each hub aggregates EVs that interface bidirectionally with the grid through smart inverters. The active and reactive power setpoints $(P, Q)$ at each hub can be controlled by an agent. In single-hub scenarios, one V2G hub is controlled for regulation, while multi-hub scenarios coordinate multiple hubs simultaneously. The RL framework includes: (i) OpenDSS-based power flow environment, (ii) RL agent for optimal hub-level control decisions, and (iii) EV fleet layer translating control outputs into physically realizable power injections subject to SOC, SOH, and inverter efficiency constraints.

The distribution network is simulated using OpenDSS where each V2G hub ($h$) is modeled as a controllable three-phase generator. The apparent power at the hub is represented as $S^{(h)} = P^{(h)} + jQ^{(h)}$, subject to active and reactive power limits $|P^{(h)}| \le P_{\max}^{(h)}$ and $|Q^{(h)}| \le Q_{\max}^{(h)}$.

\subsection{EV Fleet Model and Operational Constraints}

Each V2G hub aggregates a fleet of EVs whose individual states collectively determine available charging and discharging capability. The fleet model integrates electrochemical limits, inverter efficiency, and participation constraints to ensure physically consistent operation.

\subsubsection{Battery Power Constraints}
Each EV ($e$) is characterized by its battery capacity ($C_e$), state of charge ($SOC_e$), and state of health ($SOH_e$). Each EV's instantaneous power capability $P_{e}^{\text{bat}}$ is estimated based on its predicted operating conditions:
\begin{equation}
P_{e}^{\text{bat}} = \min(P_{\text{voltage},e}, P_{\text{current},e})
\end{equation}
%\JZ{Are $P_{e}^{\max}$ and $P_{e}^{\text{bat}}$ same here? typo?}
where $P_{\text{voltage},e}$ and $P_{\text{current},e}$ represent voltage-limited and current-limited power based on SOC and SOH. The power constraints are enforced through C-rate limitations that vary with SOC and SOH, and inverter constraints, ensuring realistic battery behavior during grid service operations.

\subsubsection{SOC and SOH Evolution}
Battery states evolve based on current throughput and degradation factors:
\begin{equation}
\begin{aligned}
SOC_e(t+\Delta t) &= SOC_e(t) + \frac{I_{e}^{\text{bat}} \Delta t}{C_e SOH_e}, \\
SOH_e(t+\Delta t) &= SOH_e(t) - \Delta SOH_{\text{cycle}} - \Delta SOH_{\text{cal}},
\end{aligned}
\end{equation}
where $I_{e}^{\text{bat}} = P_{e}^{\text{bat}} / V_e$ is the battery current (positive for charging, negative for discharging), $V_e$ is the battery voltage in V, %where $P_{\text{bat},e}$ is positive for charging and negative for discharging, 
and degradation depends on throughput energy, temperature, and duty cycle. $\Delta SOH_{\text{cycle}}$ represents cycle-based degradation from energy throughput and $\Delta SOH_{\text{cal}}$ represents calendar aging degradation from time and temperature effects.
% \RJ{explicitly mention $\Delta h_{\text{cycle}}$ and $\Delta h_{\text{cal}}$}

%At the fleet level, the hub aggregates individual SOH states to determine an effective hub SOH metric that prioritizes vehicles with higher remaining health for V2G participation.This aggregation ensures that less degraded batteries (typically offering higher driving ranges) are utilized first for grid support, while aged EVs retain capacity for their primary transportation tasks. The aggregate hub SOH, therefore, reflects a weighted measure of individual EV SOH values arranged in descending order, enabling dynamic and equitable participation control across the fleet. Furthermore, The fleet management module integrates per-vehicle power predictions to compute the available aggregate hub power $P^{(h)}_{\text{avail}}$, which constrains the hub-level power requests generated by the RL controller during fleet dispatch.

\subsubsection{Fleet-Level Power Allocation}
At the fleet level, the grid power injection required for voltage regulation is translated to behind-the-meter fleet demand using apparent power and inverter efficiency considerations. The total apparent power requested for regulation at the hub is:
\begin{equation}
S_{\text{req}}^{(h)} = \sqrt{{P_{\text{req}}^{(h)}}^2 + {Q_{\text{req}}^{(h)}}^2}
\end{equation}
where $P_{\text{req}}$ and $Q_{\text{req}}$ denote the active and reactive power injection required at the hub for voltage regulation, respectively. The battery power required to meet this requested demand is:
\begin{equation}
P_{\text{fleet}}^{(h)} = \frac{S_{\text{req}}^{(h)}}{\eta_{\text{inv}}}
\end{equation}
where $\eta_{\text{inv}}$ is the inverter efficiency. When the requested power exceeds fleet capability $P_{\text{avail}}^{(h)}$, a proportional scaling ensures feasibility:
\begin{equation}
\rho^{(h)} = \min\left(1, \frac{P_{\text{avail}}^{(h)}}{P_{\text{fleet}}^{(h)}}\right),
\end{equation}
%\HDK{$P_{\text{avail}}^{(h)}$ needs to be defined -- used in the previous equation.}
and the actual power delivered to the grid is:
\begin{equation}
P_{\text{sup}}^{(h)} = \rho^{(h)} P_{\text{fleet}}^{(h)}, \quad 
Q_{\text{sup}}^{(h)} = \rho^{(h)} Q_{\text{fleet}}^{(h)}.
\end{equation}
% \HDK{$Q_{\text{fleet}}^{(h)}$ needs to be defined.}
This hierarchical allocation mechanism ensures that fleet-level power injections remain physically consistent with individual EV capabilities, inverter limits, and SOH-dependent availability.

\subsection{Reinforcement Learning Framework}
\label{sec:rl_framework}

The voltage regulation task is formulated as a Markov Decision Process (MDP) defined by $(\mathcal{S}, \mathcal{A}, \mathcal{P}, r)$. The OpenDSS environment serves as the simulation environment, encapsulating network equations, voltage constraints, and dynamic interactions between load conditions and V2G power injections.

\subsubsection{State and Action Spaces}
The state space $\mathcal{S}$ consists of: (i) bus voltage magnitudes $V_i^{\text{pu}}$ for all monitored buses in per unit (p.u.). The action space $\mathcal{A}$ contains continuous power scaling factors for hub injections, where the agent's action vector is $\mathbf{a} = [a_P^{(1)}, a_Q^{(1)}, \ldots, a_P^{(h)}, a_Q^{(h)}]$ for $h$ hubs, which are normalized control factors bounded in $[-1,1]$.

These actions are converted to real and reactive power setpoints:
\begin{equation}
P_{\text{req}}^{(h)} = a_P^{(h)} P_{\max}, \quad 
Q_{\text{req}}^{(h)} = a_Q^{(h)} Q_{\max},
\end{equation}

%\RJ{we have action $a_P^{(h)}$ how is that converted to $P_{\text{req}}^{(h)}$ needs to be discussed. This term $\alpha_P^{(h)}$ is not connecting well. There seems to be something missing here.} where $a_P^{(h)}$ and $a_Q^{(h)}$ are normalized control factors bounded in $[-1,1]$.

\subsubsection{Reward Function}
The reward function guides the RL model to regulate voltage in the network through the V2G control:
\begin{equation}
r_t = R_{\text{vb}} - R_{\text{vp}}
\end{equation}
where:
\begin{align}
R_{\text{vb}} &= \begin{cases}
10.0 & \text{if all voltages in range} \\
0 & \text{otherwise}
\end{cases}\\
R_{\text{vp}} &= \sum_{i} \begin{cases}
(V_{\min} - V_i) \times 100 & \text{if } V_i < V_{\min} \\
(V_i - V_{\max}) \times 100 & \text{if } V_i > V_{\max} \\
0 & \text{otherwise}
\end{cases} 
\end{align}
%Here, $R_{\text{vb}}$ represents the voltage bonus reward for maintaining all voltages within acceptable limits, and $R_{\text{vp}}$ denotes the voltage penalty for voltage violations. 
The voltage limits are the traditional thresholds considered in power systems, i.e., $V_{\min} = 0.95$ p.u. and $V_{\max} = 1.05$ p.u.

%\RJ{In RL the objective is reward maximization. Depending on that any penalty is -ve and incentive is +ve. Can you double check that and change (7)?}

%\RJ{Convergence plot is missing}

\subsubsection{SAC Algorithm}
The agent employs the Soft Actor-Critic (SAC) algorithm, an entropy-regularized actor-critic method well-suited for continuous control problems. SAC maximizes the expected cumulative reward augmented with an entropy term to promote exploration:
\begin{equation}
J(\pi) = \mathbb{E}_{(s_t,a_t)\sim\mathcal{D}} 
\left[ Q_\theta(s_t,a_t) - \alpha \log \pi_\phi(a_t|s_t) \right],
\end{equation}
Here, $\pi_\phi(a_t|s_t)$ denotes the stochastic actor (policy) network parameterized by $\phi$, which outputs a probability distribution over actions given state $s_t$, and $Q_\theta(s_t,a_t)$ represents the critic network parameterized by $\theta$, which estimates the expected return for each state–action pair. The temperature coefficient $\alpha$ balances reward maximization and policy stochasticity, controlling the trade-off between exploitation and exploration. The replay buffer $\mathcal{D}$ stores past transitions $(s_t,a_t,r_t,s_{t+1})$ used for off-policy learning. %Episodes terminate after a fixed number of steps or upon convergence failure.
% \HDK{$Q_\theta$ and $\pi_\phi$ can be mentioned in the explanation below.}

\subsection{Training and Deployment Workflow}

\begin{figure}[htb!]
\centering
\includegraphics[width=\columnwidth]{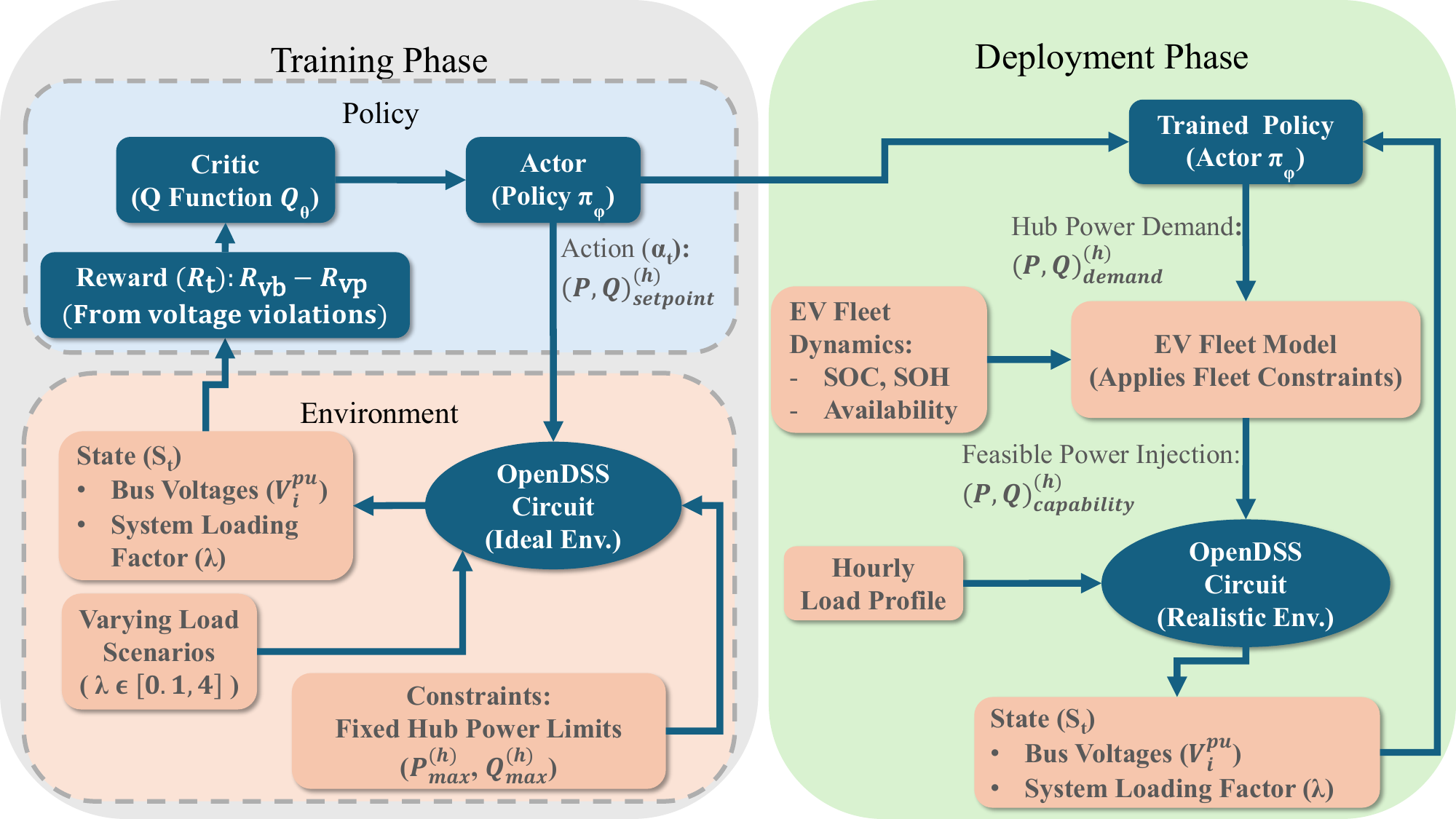}
\caption{Learning architecture and training-deployment workflow for V2G voltage regulation control.} %\JZ{make the arrow line a little bit thicker. Use ``Hub Power Demand", ``Feasible Power Injection", ``Applies Fleet Constraints", ``Hourly Load Profile"}}
\label{fig:learning_overview}
\end{figure}

The methodology employs a two-phase workflow shown in Fig.~\ref{fig:learning_overview}. Phase 1 trains the agent in an idealized environment with fixed hub power limits and no explicit fleet constraints, using time-varying load conditions through load multipliers $\lambda \in [\lambda_{\min}, \lambda_{\max}]$ applied to base loads: $P^{\text{load}}_i(\lambda) = \lambda P^{\text{base}}_i$, $Q^{\text{load}}_i(\lambda) = \lambda Q^{\text{base}}_i$. Phase 2 evaluates the trained policy with detailed fleet model enabled, where hub-level scaling ratio $\rho^{(h)}$ adjusts agent outputs based on real-time fleet availability while SOC and SOH states evolve dynamically. This two-phase design implements policy learning and physical EV constraints enforcement in different modules, ensuring operational feasibility and training stability.

\section{Case Study}
\label{sec:setup}
\subsection{Simulation Environment}
The proposed framework is validated on the IEEE 34-bus radial distribution feeder, modeled in OpenDSS and interfaced with the Gymnasium RL framework via OpenDSSDirect \cite{towers2024gymnasium,Meira2024OpenDSSDirect}. The feeder operates at a nominal voltage of 24.9 kV, with base loading of 1.8 MW and 1 MVAr. %Voltage regulators and capacitor banks are kept fixed during training to isolate the impact of the V2G hubs on voltage regulation.

\begin{figure}[thb!]    
\centering
\vspace{-10pt}
        \includegraphics[width=0.45\textwidth]{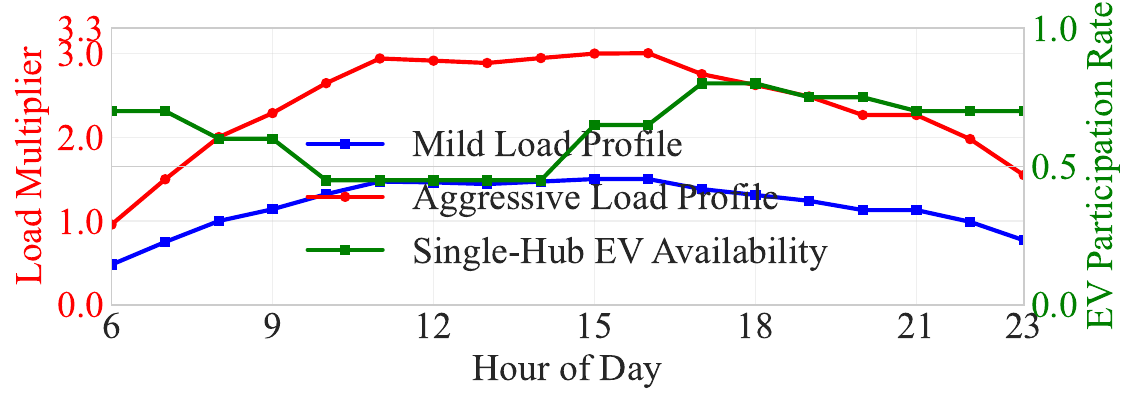}
       % \caption{}%Load Profile}
    \caption{{\textbf{Evaluation Profiles.} The left axis displays the time-varying load multipliers for both the \textbf{Mild} (peak $\lambda=1.5$) and \textbf{Aggressive} (peak $\lambda=3.0$) scenarios. The right axis indicates the corresponding EV fleet participation rate during the active V2G window (06:00--23:00).}}
   \label{fig:casestudyprofile}
\end{figure}

In the training phase, each episode comprises 100 discrete control steps with randomly sampled load multipliers $\lambda \in [0.1, 4.0]$ to ensure exposure to diverse operating conditions. {For deployment evaluation, both mild and aggressive overloading scenarios are considered, resulting in minimum voltages of 0.907~p.u. and 0.807~p.u., respectively, in the absence of regulation.} The daily evaluation horizon uses 1-hour discrete steps, with V2G services active from 6:00 to 23:00, as shown in Fig.~\ref{fig:casestudyprofile}.

%In the training phase, each episode comprises 100 discrete control steps with randomly sampled load multipliers $\lambda \in [0.1, 4.0]$, ensuring exposure to diverse operating conditions. For deployment evaluation, we considered  {both mild and }aggressive overloading scenarios, which results in the network voltage minimum to be  {0.907 p.u and} 0.807 p.u, respectively. without regulation. The daily evaluation horizon uses 1-hour discrete steps. The V2G service window spans from 6:00 to 23:00. V2G services are not active during hours 0–5 (midnight to 5:59 AM) when load conditions are light and EVs are typically charging to prepare for daily use. As a result we focus on the loading condition in this window as shown in Fig.~\ref{fig:dailyprof}.

For single-hub evaluation, one V2G hub is connected at bus 890. For multi-hub coordination, five V2G hubs are connected at buses 890, 844, 832, 830, and 860. Hubs have rated power of 500 kW (active) and 400 kVAr (reactive) each. Each EV has a 75 kWh battery, initial SOC range of 0.2 to 0.9, SOH of 0.95, charging/discharging efficiency of 0.96, and C-rate limit of 0.5C. The single-hub scenario represents a single fleet hub with realistic EV availability patterns, while the multi-hub scenario represents coordination across multiple independent fleet hubs where the aggregator coordinates power dispatch across all hubs.

For the single-hub case study, time-varying EV availability schedules are incorporated to model realistic fleet participation patterns, where the number of available EVs varies hourly based on delivery schedules (45-85\% availability). For the multi-hub case study, the evaluation focuses on coordination capabilities by assuming sufficient EV capacity at each hub, allowing the analysis to isolate the benefits of multi-hub coordination from availability constraints.

\subsection{Control Algorithms}

\subsubsection{RL Agent}
{The control agent utilizes the Soft Actor-Critic (SAC) algorithm. The actor and critic networks each consist of two hidden layers with 256 ReLU units. Key hyperparameters include a learning rate of $3 \times 10^{-4}$, batch size of 256, discount factor $\gamma=0.99$, and replay buffer size of $10^6$.}

\subsubsection{Baseline Strategy}
 {To validate performance, we benchmark against a decentralized Droop control strategy. Each hub implements autonomous piecewise-linear Volt-Var ($Q(v)$) and Volt-Watt ($P(v)$) curves with a deadband of $\pm 0.02$ p.u. and saturation at $[0.90, 1.10]$ p.u. This represents a state-of-the-art local control baseline where inverters act independently without inter-hub communication.}

\section{Results}
\label{sec:results}

%%%%%%%%%%%%%%%%%%%%%%%
\subsection{Single-Hub Voltage Regulation Performance}
 {This section evaluates the voltage regulation capability of a single V2G hub under mild and aggressive loading conditions, with the objective of distinguishing controller capability from realistic deployment limits. To this end, we first report results without EV constraints to isolate how effectively RL and droop control can support voltage when sufficient energy is available, and then enforce EV constraints to quantify how availability and SOC limitations reduce achievable performance. Mild loading is presented first as a representative undervoltage case, followed by aggressive loading to illustrate how single-hub performance degrades under extreme stress.}

\setlength{\tabcolsep}{2.6pt}
\begin{table}[!t]
\centering
\caption{Single-hub voltage regulation summary. Statistics are computed from hourly mean bus voltages across the feeder; violation hours count hours with mean voltage $<0.95$~p.u.}
\label{tab:single_hub_summary}
\begin{tabular}{lcccccccc}
\toprule
& \multicolumn{4}{c}{Mild loading} & \multicolumn{4}{c}{Aggressive loading} \\
\cmidrule(lr){2-5} \cmidrule(lr){6-9}
Scenario
& Mean & Min & Max & Viol.
& Mean & Min & Max & Viol. \\
\midrule
Baseline
& 0.963 & 0.907 & 1.048 & 13
& 0.883 & 0.807 & 1.017 & 17 \\
RL (no EV)
& 0.989 & 0.942 & 1.048 & 6
& 0.896 & 0.812 & 1.017 & 17 \\
Droop (no EV)
& 0.987 & 0.943 & 1.048 & 6
& 0.898 & 0.821 & 1.017 & 17 \\
RL (EV-constr.)
& 0.969 & 0.910 & 1.048 & 12
& 0.888 & 0.810 & 1.017 & 17 \\
Droop (EV-constr.)
& 0.972 & 0.908 & 1.048 & 11
& 0.889 & 0.807 & 1.017 & 17 \\
\bottomrule
\end{tabular}
\end{table}
 {Table~\ref{tab:single_hub_summary} summarizes the single-hub voltage regulation results across both loading profiles. Under mild loading and without EV constraints, both RL and droop substantially improve voltage conditions relative to the baseline, reducing violation hours from 13 to 6 and achieving similar minimum voltages. When EV constraints are enforced, the achievable gains are significantly reduced for both controllers, with violation hours remaining close to the baseline level. In this constrained setting, RL and droop exhibit comparable performance, indicating that fleet availability and SOC limits, rather than the specific control strategy, dominate single-hub voltage regulation outcomes. These results suggest that, for a single hub, fleet availability rather than inverter rating becomes the primary performance bottleneck once realistic constraints are applied. Under aggressive loading, voltage violations persist across all single-hub configurations, regardless of control strategy or constraint setting. While both RL and droop provide marginal improvements in mean and minimum voltage relative to the baseline in the unconstrained case, enforcing EV constraints eliminates most of these gains, with violation hours remaining unchanged at 17.} %This highlights the fundamental limitation of single-point voltage support under severe feeder-wide stress, where localized injections are insufficient to sustain meaningful voltage recovery.
 
\begin{figure}[thb!]
    \centering
    \begin{subfigure}[b]{0.5\textwidth}
        \includegraphics[width=\textwidth]{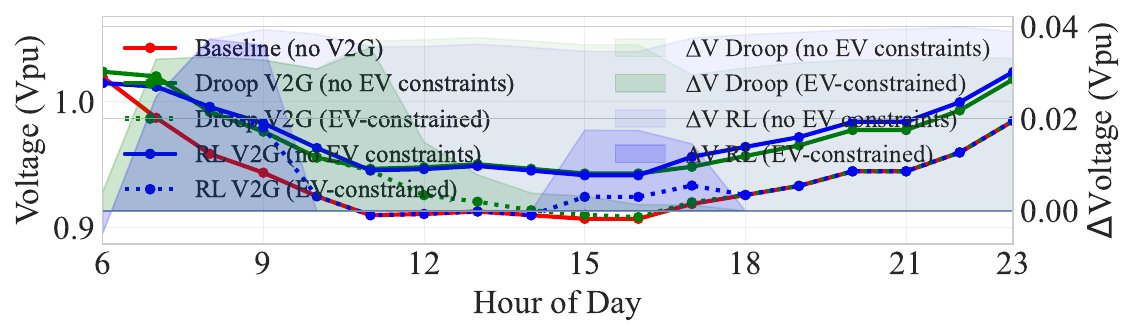}
        \caption{Mild loading}
        % \caption{Mild loading voltage profile and mean voltage improvement relative to baseline}
        \label{fig:single_mild}
    \end{subfigure}
    \begin{subfigure}[b]{0.5\textwidth}
        \includegraphics[width=\textwidth]{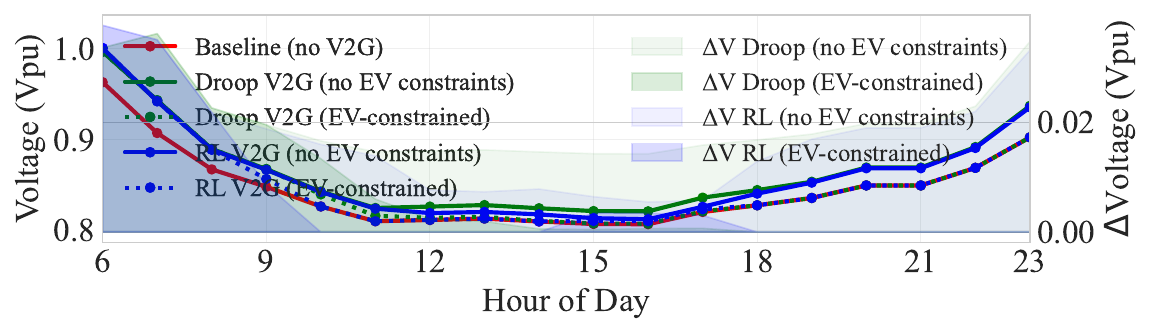}
        \caption{Aggressive loading}
        % \caption{Aggressive loading voltage profile and mean voltage improvement relative to baseline}
    \end{subfigure}
    \begin{subfigure}[b]{0.45\textwidth}
        \includegraphics[width=\textwidth]{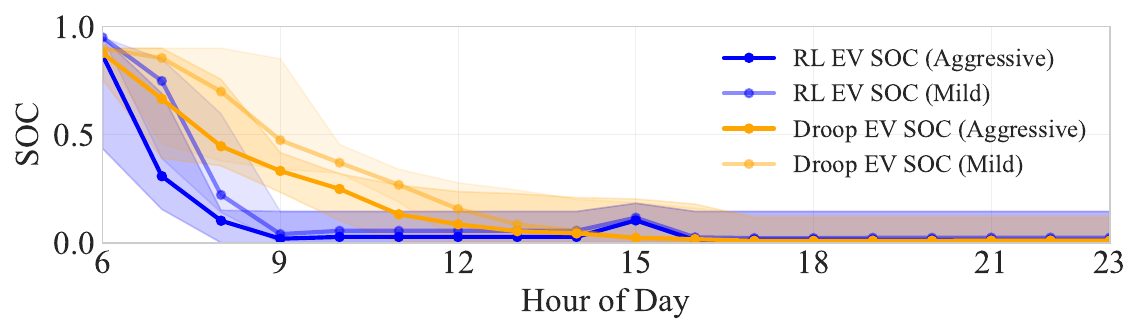}
        \caption{Fleet average SOC}
    \end{subfigure}
    \begin{subfigure}[b]{0.45\textwidth}
        \includegraphics[width=\textwidth]{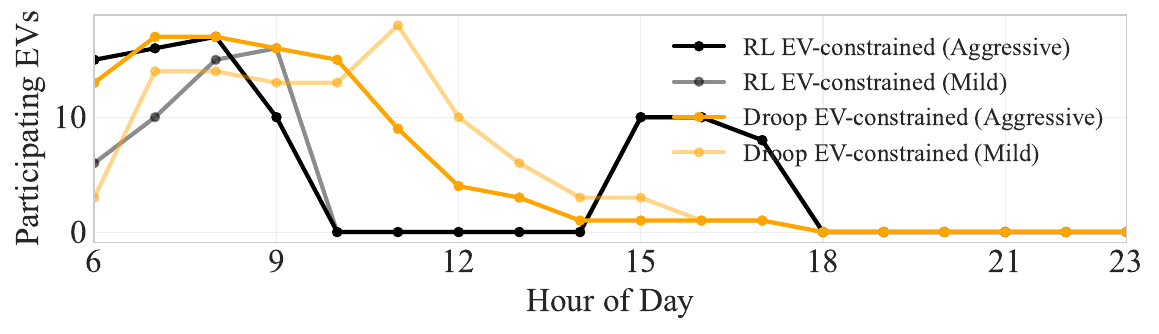}
        \caption{Number of participating EVs}
    \end{subfigure}
    \caption{ {\textbf{Single-Hub Voltage Regulation Performance.} }
    (a)–(b) Mean feeder voltage profiles and improvements under mild and aggressive loading. 
    (c) Fleet-average SOC. 
    (d) Participating EV count. 
    Voltages are averaged across all buses.}
    %\JZ{increase the font size in the figure.}}
    \label{fig:single_hub_aggressive}
    %\RJ{Is this Vmin? If so specify in (a) and (b) this is the minimum network voltage because the voltage regulation effects may be pronounced at buses surrounding hubs.}
\end{figure}

 {Figure~\ref{fig:single_hub_aggressive} provides time-series context for these results. The voltage improvements concentrate during periods when EV availability and SOC headroom are highest, and diminish as participation decreases or batteries approach lower SOC levels. In the aggressive case, deeper SOC drawdown and tighter availability further constrain support, consistent with the limited improvements observed in Table~\ref{tab:single_hub_summary}. Overall, these results demonstrate that while single-hub V2G control can improve voltage margins under moderate conditions, its effectiveness is fundamentally constrained by topology and availability, motivating the need for coordinated multi-hub support.}

% Fig.~\ref{fig:single_hub_aggressive} shows that voltage support tracks EV availability and SOC headroom, with aggressive loading exhibiting deeper drawdown and reduced effectiveness.

%%%%%%%%%%%%%%%%%%%%%%%

\subsection{Multi-Hub Coordination}

 {We evaluate coordinated control across five hubs under mild and aggressive loading to benchmark against droop and to assess generalization and stress response. The mild case reflects typical undervoltage operation, while the aggressive case stresses voltage recovery when feeder-wide support is required.}

\setlength{\tabcolsep}{2.85pt}
\begin{table}[!t]
\centering
\caption{Multi-hub voltage regulation summary. Statistics are computed from hourly mean bus voltages; violation hours count hours with mean voltage $<0.95$~p.u.}
\label{tab:multi_hub_summary}
\begin{tabular}{lcccccccc}
\toprule
& \multicolumn{4}{c}{Mild loading} & \multicolumn{4}{c}{Aggressive loading} \\
\cmidrule(lr){2-5} \cmidrule(lr){6-9}
Scenario
& Mean & Min & Max & Viol.
& Mean & Min & Max & Viol. \\
\midrule
Baseline
& 0.963 & 0.907 & 1.048 & 13
& 0.883 & 0.807 & 1.017 & 17 \\
RL (coord.)
& 1.025 & 1.007 & 1.048 & 0
& 0.949 & 0.903 & 1.017 & 15 \\
Droop (coord.)
& 1.024 & 1.004 & 1.048 & 0
& 0.998 & 0.940 & 1.035 & 2 \\
\bottomrule
\end{tabular}
\end{table}

\begin{figure}[thb!]
    \centering
    \begin{subfigure}[b]{0.45\textwidth}
        \includegraphics[width=\textwidth]{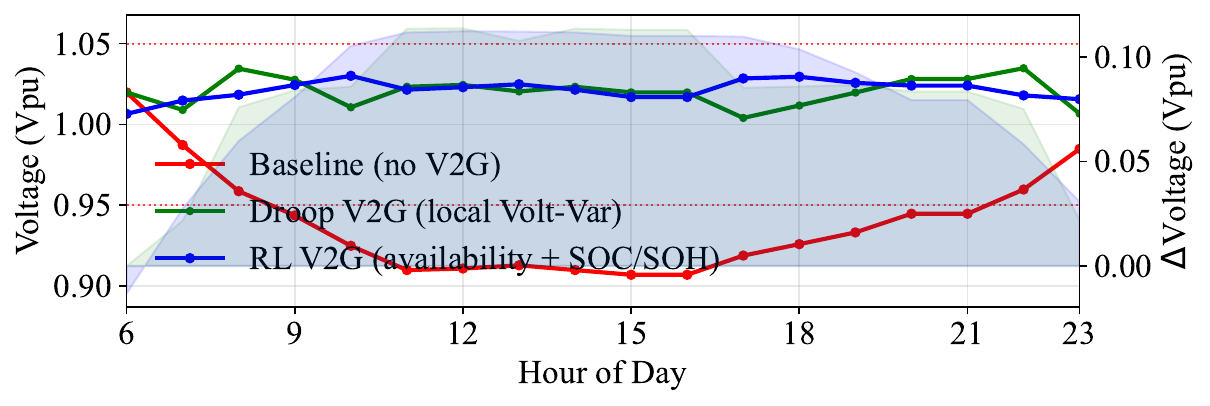}
        \caption{Mild loading}
        \label{fig:multi_mild}
    \end{subfigure}
    \begin{subfigure}[b]{0.45\textwidth}
        \includegraphics[width=\textwidth]{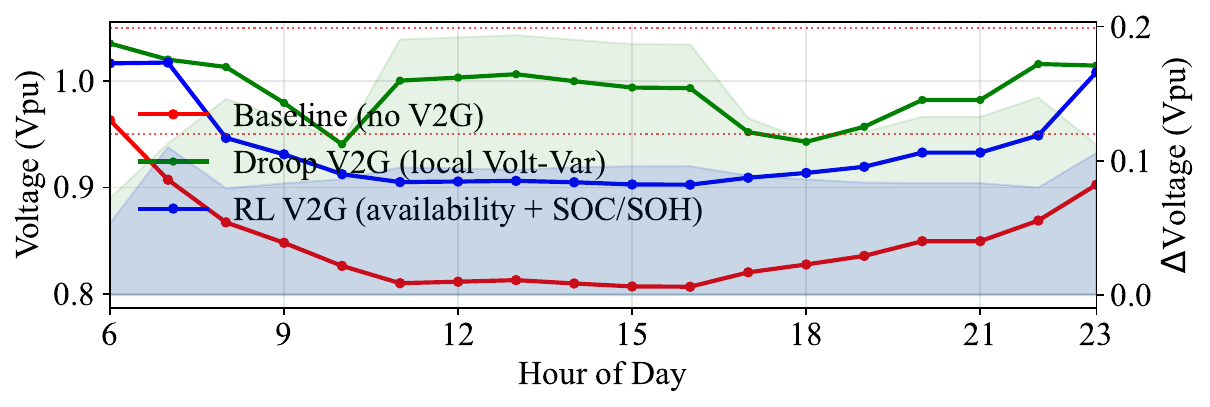}
        \caption{Aggressive loading}
        \label{fig:multi_aggr}
    \end{subfigure}
    % \caption{\textbf{Multi-Hub Voltage Regulation.} 
    % (a) Mild loading: coordinated control maintains feeder voltages near nominal relative to the baseline. 
    % (b) Aggressive loading: coordinated control provides feeder-wide voltage support under severe loading conditions. 
    % Subplots (a) and (b) show mean voltages across all network buses, illustrating the average voltage response under coordinated V2G control.}
    \caption{\textbf{Multi-Hub Voltage Regulation.} 
    (a) Mild loading and (b) aggressive loading scenarios under coordinated V2G control. 
    Subplots show mean voltages averaged across all network buses, illustrating feeder-wide voltage response relative to the baseline.}

%\JZ{increase the font size in the figure.}}
%\RJ{here also (a) and (b) caption specify minimum network voltage}
    \label{fig:multi_hub_aggressive}
\end{figure}

% In the mild scenario, both RL and droop eliminate voltage violations and achieve near-identical mean/minimum voltages, indicating that mild conditions can be handled by either strategy. In the aggressive scenario, both controllers improve the feeder-wide profile relative to baseline; droop attains higher mean/minimum voltage and fewer violation hours, while RL still provides a substantial uplift over baseline. Peak total active support reaches 562.6~kW (mild) and 905.0~kW (aggressive), and is distributed across hubs according to local voltage conditions and hub capacity.

% In the aggressive scenario (Fig.~\ref{fig:multi_hub_aggressive}), the baseline exhibits sustained sag. Both controllers lift the feeder-wide profile relative to baseline, but droop achieves a higher mean and minimum voltage and fewer violation hours in this case (Table~\ref{tab:multi_hub_aggressive_summary}). The coordinated RL policy still provides substantial improvement over baseline, indicating that spatial coordination remains critical when single-point support is insufficient.

 {Table~\ref{tab:multi_hub_summary} summarizes the coordinated multi-hub results, with time-series behavior illustrated in Fig.~\ref{fig:multi_mild} and Fig.~\ref{fig:multi_aggr}. }

 {Under mild loading, both coordinated RL and droop control eliminate voltage violations and achieve nearly identical mean and minimum voltages, indicating that moderate voltage deviations can be effectively handled by either strategy. The RL policy maintains voltages close to nominal without introducing over-voltage risks, suggesting that it captures the underlying voltage sensitivity of the network rather than overfitting to a specific loading trajectory.}

 {Under aggressive loading, the uncontrolled baseline exhibits sustained feeder-wide voltage sags, while both coordinated controllers substantially improve the voltage profile, confirming the importance of spatial coordination across multiple hubs when single-point support is insufficient. In this regime, the droop controller attains higher mean and minimum voltages and reduces violation hours more effectively by aggressively driving inverter outputs toward their limits based on local voltage measurements. The RL controller does not match droop performance in this stress case, but still delivers a consistent and nontrivial voltage uplift across the feeder through coordinated multi-hub dispatch.} %Across both scenarios, peak total active power support reaches 562.6~kW under mild loading and 905.0~kW under aggressive loading, with injections distributed across hubs according to local voltage conditions and hub capacity. 
 {Overall, while rule-based droop remains highly effective for saturation-driven voltage correction under extreme conditions, the coordinated RL framework demonstrates a flexible and extensible approach for learning feeder-wide coordination, providing a promising foundation for future integration of fleet-aware constraints and additional system-level objectives.}

% Table~\ref{tab:voltage_stats} summarizes the voltage performance across different scenarios. The baseline system experiences significant voltage violations, particularly under aggressive loading conditions where during 17 hours of the day it exceeds the ±5\% voltage band. The V2G controllers demonstrate varying effectiveness: under aggressive loading, multi-hub systems reduce violations from 17 to 15 hours and achieve substantial voltage improvements (mean 0.949 p.u. vs. baseline 0.883 p.u.), while single-hub systems show limited improvement (0.888 p.u., 17 violation hours unchanged). Multi-hub systems demonstrate 3-6x better average voltage improvements compared to single-hub configurations.

\section{Conclusion}
\label{sec:conclusion}

This paper presents a V2G coordination framework for voltage regulation using reinforcement learning, demonstrating both single-hub and multi-hub coordination capabilities. The framework introduces a two-phase operational approach with policy learning based on charger hub integrated distribution feeder model and deployment module that enforces the realistic fleet constraints, facilitating training stability while ensuring practical feasibility.

%The multi-hub configuration's superior performance suggests that distributed V2G systems offer significant advantages for large-scale voltage regulation applications, particularly under aggressive loading scenarios where single-hub capacity may be insufficient to achieve desired regulation. Future research directions include extending the framework to larger distribution networks, incorporating multi-agent control, and investigating the impact of vehicle travel and logistics on voltage regulation.

 {The multi-hub results demonstrate that coordinated RL can provide meaningful feeder-wide support, but a local droop baseline can still outperform it under aggressive stress. This motivates future work on constraint-aware objectives, including battery-degradation-aware optimization, and extensions to larger feeders, multi-agent coordination, and the integration of vehicle travel and logistics constraints.}

\section*{Acknowledgment}%\RJ{Please check grants. I used same as KPEC}
This work was partially supported by the Department of the Navy, Office of Naval Research under ONR award number N00014-21-1-2530 and National Science Foundation under Award 2229417. 

%\JZ{please cite Roshini's Nature Comms paper. I don't see any refs published in IEEE Trans. It is better to cite 1-2 papers from IEEE Trans. on Power Systems or Smart Grid if any.}

\bibliographystyle{IEEEtran}
\bibliography{GMPES}

\end{document}